\documentclass[pdflatex,sn-mathphys]{sn-jnl}
\usepackage{longtable}
\usepackage{graphicx}
\usepackage{booktabs}
\usepackage{multirow}
\usepackage{enumitem}
\begin{document}

\title{Machine Learning Simulates Agent-Based Model Towards Policy}

\author[1,3]{\fnm{Bernardo} \sur{Alves Furtado}}\email{bernardo.furtado@ipea.gov.br}
\author[1,2]{\fnm{Gustavo Onofre} \sur{Andreão}}\email{gustavo.93.andreao@gmail.com}

\affil[1]{\orgdiv{Department of Sectorial, Infrastructure and Innovation Studies}, \orgname{Institute for Applied Economic Research}}

\affil[2]{\orgdiv{Institute of Economics}, \orgname{UNICAMP}}

\affil[3]{\orgname{National Council for Scientific and Technological Development}} 

\abstract{Public Policies are not intrinsically positive or negative. Rather, policies provide varying levels of effects across different recipients. Methodologically, computational modeling enables the application of multiple influences on empirical data, thus allowing for heterogeneous response to policies. We use a random forest machine learning algorithm to emulate an agent-based model (ABM) and evaluate competing policies across 46 Metropolitan Regions (MRs) in Brazil. In doing so, we use input parameters and output indicators of 11,076 actual simulation runs and one million emulated runs. As a result, we obtain the optimal (and non-optimal) performance of each region over the policies. Optimum is defined as a combination of GDP production and the Gini coefficient inequality indicator for the full ensemble of Metropolitan Regions. Results suggest that MRs already have embedded structures that favor optimal or non-optimal results, but they also illustrate which policy is more beneficial to each place. In addition to providing MR-specific policies' results, the use of machine learning to simulate an ABM reduces the computational burden, whereas allowing for a much larger variation among model parameters. The coherence of results within the context of larger uncertainty--\textit{vis-à-vis} those of the original ABM--reinforces robustness of the model. At the same time the exercise indicates which parameters should policymakers intervene on, in order to work towards precise policy optimal instruments.}



\maketitle
\section{Computational Models for Policy}\label{sec1}

Policymakers know that policies, as broad as they may be, lead to different results in different contexts \cite{stone_understanding_2017, Mitchell.Woodman2010Regulationandsustainableenergysystems, Gawel.etal2016Supportpoliciesforrenewablesinstrumentchoiceandinstrumentchangefromapublicchoiceperspective}. Why certain policies impact different regions differently, however, is difficult to pinpoint \cite{cravo_economic_2013, Boschma2013ConstructingRegionalAdvantageandSmartSpecializationComparisonofTwoEuropeanPolicyConcepts, Lundvall2007NationalInnovationSystemsAnalyticalConceptandDevelopmentTool}. Interest groups, institutions and context may be relevant factors for how policies are elaborated, implemented, revised and assessed \cite{mueller_why_2020, Nelson1994Thecoevolutionoftechnologyindustrialstructureandsupportinginstitutions, Hochstetler.Kostka2015WindandSolarPowerinBrazilandChinaInterestsStateBusinessRelationsandPolicyOutcomes}. In that sense, existing structural factors, such as households attributes, income and location, along with characteristics of businesses and government might play a role \cite{Faber.etal2010ExploringdomesticmicrocogenerationintheNetherlandsAnagentbaseddemandmodelfortechnologydiffusion}.

Computational models may act as a coherent set of sequential rules and procedures that help understand, explore, and learn about effects and impacts of policies \cite{Tesfatsion2011AgentbasedModelingandInstitutionalDesign}. Models allow for the anticipation of possible results \cite{doi:10.1098/rsos.172096, deMarchi.Page2014AgentBasedModels, Faber.etal2010ExploringdomesticmicrocogenerationintheNetherlandsAnagentbaseddemandmodelfortechnologydiffusion}. Additionally, computational models make counterfactual tests not only feasible, but a must when making decisions on policies \cite{Lamperti.etal2018FarawaySoCloseCoupledClimateandEconomicDynamicsinanAgentbasedIntegratedAssessmentModel,Lespagnol.Rouchier2018TradingVolumeandPriceDistortionAnAgentBasedModelwithHeterogenousKnowledgeofFundamentals,Huang.etal2010Financialcrisesandinteractingheterogeneousagents}. Policymakers facing complex scenarios would be in a better position to make decisions when they have subsidies from an array of factual and counterfactual indicators\footnote{\cite{arthur_inductive_1994} understands that inductive reasoning plays an important part on the policymaking activity.}\cite{Brenner.Werker2009Policyadvicederivedfromsimulationmodels, Schmidt.etal2016DodeploymentpoliciespicktechnologiesbynotpickingapplicationsAsimulationofinvestmentdecisionsintechnologieswithmultipleapplications}. 

Agent-based models (ABMs) fit a class of computational models in which the emphasis lies on the agents of a system, the minimum decision-making units, and their interactions \cite{Tesfatsion2003Agentbasedcomputationaleconomicsmodelingeconomiesascomplexadaptivesystems, Arthur2006OutofEquilibriumEconomicsandAgentBasedModeling}. It is a bottom-up approach in which agents follow deterministic rules and interact to reproduce known patterns \cite{Axelrod1997Thecomplexityofcooperationagentbasedmodelsofcompetitionandcollaboration, Janssen.Ostrom2006Empiricallybasedagentbasedmodels}. Moreover, ABMs are discrete, dynamic systems in which agents of various kind and their environment co-evolve following algorithmic steps \cite{epstein_growing_1996, Vazquez.Hallack2018TheroleofregulatorylearninginenergytransitionThecaseofsolarPVinBrazil}. ABMs are tools that social scientists, in particular, may use to systematically codify tacit knowledge and behaviors and "animate" trajectories in order to experiment \cite{galan_errors_2009, Chen.etal2001Testingfornonlinearstructureinanartificialfinancialmarket, Ehrentreich2008TheOriginalSantaFeInstituteArtificialStockMarket}.

Specifically, PolicySpace2 (PS2) is an spatial-economic agent-based model that uses census and geographical data for households, businesses and municipalities within Metropolitan Regions (MRs) to simulate interactions in the housing, labor, goods and services markets. The empirical, intraurban model runs from 2010 to 2020 and is used to simulate three policy tests. Validation is made to Brasilia, the capital of Brazil, and additional results are also presented to four other MRs \cite{alves_furtado2022}. 

We use PolicySpace2 (PS2) open source as our simulation baseline model for this paper. However, whereas PS2 ran manual--hundreds--combinations of parameters for its sensitivity analysis, we design a surrogate model that expands the analyses to one million possible combinations. 

A surrogate model is a computational artifice that simultaneously mimics the procedures of a model using its inputs and outputs, whereas saving time and maximizing outputs. "An emulator is thus a model of a model: a statistical model of the simulator, which is itself a mechanistic model of the world" \cite[p.2]{paleyes_emulation_2019}.  Surrogate models provide a simple and cheap option to emulate the original run of a model to a higher order of magnitude \cite{vanderHoog2019SurrogateModellinginandofAgentBasedModelsAProspectus, tenBroeke.etal2021TheUseofSurrogateModelstoAnalyseAgentBasedModels}. 

Given this context, our research question is twofold. First, we want to learn whether different policies generate optimal results in different Metropolitan Regions? Or do they always influence the same? In other words, we would like to know whether each policy affects the different MRs heterogeneously or if there is one-size-fits-all policy for all places. Secondly, we wonder whether there are model parameters and rules that would have higher (or lower) values associated to optimal scenarios. Thus, which parameters and rules would demand (if possible) policymakers intervention in order to nudge their MRs towards optimal scenarios? 

In order to answer these questions we applied Machine Learning (ML) procedures to create a surrogate model that emulated an Agent-Based Model, PolicySpace2, and explored results of three alternative policies against the no-policy baseline for an array of 46 Brazilian MRs. As such, we anticipated empirical scenarios for each policy and each MR and contrasted these results with the absence of policy. Results expanded the test originally made by the ABM--which implied a quasi-manual test of 11 thousand runs on mostly five MRs--and amplified the combination of parameters values to one million runs. 

We chose a relational, comparative optimum ex-post target for our policy analysis. Basically, the optimal status results simultaneously imply that the Metropolitan Region has achieved the highest quartile of GDP and the lowest quartile of Gini coefficient. This choice was made to train the ML and execute the expansion of runs into a larger scope that included all parameters and MRs. Hence, optimal in this case refers to GDP and Gini coefficient indicators comparatively to all other MRs.

The contribution of the paper is to expand the empirical results of a policy-test ABM while bringing more uncertainty and more robustness to the simulation process. More uncertainty as we emulated the ABM with a much larger set of combinations among parameters that enter the model, compared to the original ABM run. More robustness as we explored results over one million outputs instead of the original 11,076 simulation runs.

Additionally, we produced a rank of Metropolitan Regions intrinsic attributes that measures their response to an specific pair of optimal policy indicator targets and policy proposal. The response also illuminated the best policy for each individual MR. Finally, we presented a surrogate model that qualify as a valid approach for spatial ABM generalization.

\section{Literature background}
In this section we discuss the pertinent literature background. First we briefly review the literature on ABM and their use for policy analysis. Then we review the literature on ML and then intersection with ABM: surrogate models.

ABMs are simulation models that highlight interactions between decision-making units, the agents, and their local environment. ABMs basically consist of a discrete system in which the states of the agents and the environment are updated according to explicit rules \cite{epstein_growing_1996}. 

Policy analysis specifically seem to be a perfect use of ABMs. First, public policies constitute complex systems with a number of ever-changing agents (citizens, politicians, businessmen, scientists, institutions) acting and reacting in time and space, with limited information, and unaware of others rationality \cite{furtado_modeling_2015,geyer_handbook_2015,arthur_inductive_1994}. 

Secondly, ABMs do not need to work with a single--likely to fail \cite{mueller_why_2020}--policy goal. Quite on the contrary, ABMs are fit to work as a communication tool \cite{gilbert_computational_2018} that makes the analysis systemic and visual, enabling parties from different backgrounds, experiences and disciplines to converse and exchange, whilst producing probabilistic scenarios. 

Thirdly, computational modeling in general and ABMs in particular are suitable for experimenting. Counterfactuals and \textit{what-if} questions are formulated and tested within a safe environment, with costs restricted to the computational burden, and the time-consumed from modelers, programmers and policymakers. 

There has been a plethora of models using ABMs in a variety of disciplines, methods and applications \cite{edmonds_simulating_2017,lee_complexities_2015,heppenstall_agent-based_2012,dawid_agent-based_2018,adepetu_agent-based_2016,scott_simdrink:_2016,ingham-dempster_tim_agent-based_2017,van_der_veen_exploring_2017}. However, a much smaller number has been actually applied and used to aid and change real policy \cite{carley_biowar_2006,kerr_covasim_2021,gilbert_computational_2018}. Mainly, the difficulty comes from the hurdle of bridging "the gap between policy practice, often expressed in qualitative and narrative terms, and the scientific realm of formal models" \cite[p.1]{ahrweiler_modelling_2015}. Nevertheless, ABMs have proved to be valuable policy tools \cite{gilbert_computational_2018}, when observing its limitations \cite{aodha_pitfalls_2017}. 

In this paper, we focus mainly on the expansion of the original model's  capacity to anticipate policy trajectories and thus enable policy decision-making. Hence, the analysis supports comparison and evaluation of alternative policies, or rather, alternative decisions on public money investing, before actually spending taxpayers' money.

\subsection{Machine learning and surrogate models}
ML consists of a series of computer methods in different areas within mathematics and statistics that attempt to automate the processes of finding, sorting and weighting different patterns in data sets. Moreover, ML aims at finding "hidden rules" in order to predict and analyze data \cite{Breiman2001Randomforests}. This is a relevant characteristic that differentiates ML from statistics or econometrics: organizing data from various sources with an emphasis on predicting and forecasting. 

ML is typically divided between supervised and unsupervised learning. Supervised learning has an optimal, target or desired output for its models. Unsupervised learning has no target and identifying patterns or clustering is the main goal. In this paper, we apply a supervised learning process in which the target is an inter-city comparative optimal duet of indicators (see subsection Methods: Procedures). 

Machine Learning also include semi-supervised and reinforcement learning. In semi-supervised ML, some data contain labels--whereas the complementing data do not--, and it is mostly used for classification. In fact, information on the labeled data may be used to inform the unlabeled ones \cite{van_engelen_survey_2020}. 
Reinforcement learning focuses on a reward function that guides the learning process of agents.\footnote{For a panorama of different algorithms and approaches, see \cite{wiering_reinforcement_2012}.} 

Random forest is a common method within supervised ML. A random forest consists of a selection of subsets that branch randomly into different trees. Those subsets are independent and non-identical decision trees defined by independent random vector parameters and input data. With a large number of trees, the algorithm then searches for the most popular class through equal-weight voting \cite{Breiman2001Randomforests, Ren.etal2017Researchonmachinelearningframeworkbasedonrandomforestalgorithm, Mishina.etal2015BoostedRandomForest}. 

In this paper we tested different ML methods and chose a random forest surrogate model that emulates the original simulation model, PS2 \cite{alves_furtado2022}. "A surrogate model is an approximation method that mimics the behavior of an expensive computation or process. (...) [a surrogate model] can be trained to represent the system, learning a mapping from simulator inputs to outputs" \cite[p.11]{lavin_simulation_2021}. Moreover, surrogates are essentially emulators that substitute an actual simulation model for a simpler method in order to produce results through less demanding means, both in relation to time and to processing power. Through the use of ML, it reiterates over the ABM using pre-processed input and output information.

A surrogate model has advantages that seem to be adequate for our proposed exercise:
\begin{enumerate}
    \item \textit{Generalization}. A surrogate enables many more combinations of input parameters. In practice, the surrogate model works as if we had run the original simulation model a much higher number of times, whilst testing a (much) wider range of possible parameters.
    \item \textit{Practicality and speed}. Running a surrogate 1,000,000 times is way faster (and actually feasible) than running 1,000 regular simulations. It is a different order of magnitude concept. The simulations are emulated, rather than actually run. This is especially valid for models with geographical information and a large number of agents.
    \item \textit{Statistical extrapolation}. Surrogate models' larger number of runs bring statistical relevance for the results and provides robustness for the extrapolation. 
\end{enumerate}

The aim of our surrogate is to extract the most information of the model with as little computational time as possible, choosing an optimal target while doing so. We revise surrogate models in relation to the three main steps in producing a surrogate.

In the ML surrogate proposed by \cite{Lamperti.etal2018Agentbasedmodelcalibrationusingmachinelearningsurrogates}, the authors explore the parameter space through the use of a non-parametric ML surrogate without prior knowledge of the spatial distribution of data. Starting with a sample of initial conditions values (through Sobol sampling), random subsets are drawn, ran through the ABM and then evaluated by user-defined criterion of positive or negative. Then the surrogate is learned through a ML algorithm using a variation of random forest. By repeating the steps, the surrogate predicts the probability of false negatives (non-optimal results that are actually optimal) and is developed focusing on producing less false negatives. In a nutshell, the ABM is ran each time, it processes the data, labels and evaluates the results.

The surrogate of \cite{Lamperti.etal2018Agentbasedmodelcalibrationusingmachinelearningsurrogates} is directly comparable to the surrogate models that use Bayesian emulation or kriging, such as the model of \cite{tenBroeke.etal2021TheUseofSurrogateModelstoAnalyseAgentBasedModels}. According to the authors, their kriging surrogate is built through: model specification through a prior-distribution of hyper-parameters, then the ABM is ran and results are gathered by the use of an integration function of the generalized least squares estimator of positive-definite symmetric matrices. The modeler then must choose a simulator: a multi-output (MO) emulator or a many to single-output. The authors choose the multi-output (MO) emulator due to convenience and least computational time \cite{tenBroeke.etal2021TheUseofSurrogateModelstoAnalyseAgentBasedModels}. By using the MO emulator, one may use the produced surrogate to make model predictions. Their surrogate is quite similar to that of \cite{Conti.OHagan2010Bayesianemulationofcomplexmultioutputanddynamiccomputermodels}, especially in relation to the discussion of multi-output emulators versus other options. Essentially, their surrogate separates data into training and testing classifications, expanding the training data until sufficient performance (decided \textit{ad hoc} by the modeler) is reached.

The surrogate of \cite{Hayashi.etal2016ImprovingBehaviorPredictionAccuracybyUsingMachineLearningforAgentBasedSimulation} focuses on the parameter space of the simulation. The authors use four ML methods, with neural networks and random forests being part of those tested. The authors analyze many inputs to one output, again separating between "positive" and "negative" outcomes. ML feeds back the ABM by revising the inner workings of agents with each round of the simulation restarting with initial conditions given by the ML process. Their models connects the ABM and ML portions by comparing their results, attempting to minimize the error between the outputs of the ABM and the ML surrogate. 

The surrogate of \cite{Edali.Yucel2019Exploringthebehaviorspaceofagentbasedsimulationmodelsusingrandomforestmetamodelsandsequentialsampling} relies heavily on random forest and focuses on output predictions. For the training and fitting of the surrogate, they use the latin hypercube sampling and then sequential sampling. It combines random forests with uncertainty sampling in order to train the surrogate model.

The surrogates we reviewed interact with an ABM model or the modeler itself to increase the quality of the prediction or the performance of the original model. Our proposal is somewhat independent of the original ABM. Our surrogate uses input and output from the original ABM to learn the inner mechanisms--the mapping--that transforms inputs into outputs. Once trained, a new analysis that takes advantage of a much larger parameter space (unfeasible for the case of the original ABM) is applied with a clear target of policy optimality. The results are then compared providing a counterfactual panorama for policymakers. 

In the next section, we describe first the basics of the original PS2 model, then the data generated by PS2 and used as input to our surrogate model. After that, we detail the procedures and steps that we use in running the emulator.

\section{Methods}
\subsection{Baseline Agent-Based Model: PolicySpace2 (PS2)}

PolicySpace2 (PS2) is an open source, readily available\footnote{\url{https://github.com/BAFurtado/policyspace2}.} ABM that focuses on three alternative policy-schemes for the case of Brazilian MRs. According to the authors, PS2 is defined "as a primarily endogenous computational agent-based model (ABM) that includes mortgage loans, housing construction, taxes collection and investments, with firms and households interacting in real estate, goods and services, and labor markets" \cite[par.1.5]{alves_furtado2022}. The default run of PS2 is for the MR of Brasília from 2010 to 2020, using census and business data, along with detailed geographical intraurban (census tract equivalent) boundaries. 

The main goal of PS2 is to contrast and compare three competing policy investments alternatives for the case of Brasília. Given low-income households, municipalities may either (a) acquire houses to transfer (Property acquisition); (b) provide rental vouchers (Rent vouchers); or (c) make monetary transfers (Monetary aid). These policies are tested against a no-policy baseline and results are compared using macroeconomic indicators and their trajectories.

The results of PS2 consistently show that Rent vouchers (b) and Monetary aid (c) seem to be the most beneficial policy choice producing higher GDP with a lower Gini coefficient. These two options also reduced the percentage of households defaulting on rent and going any month without goods and services consumption. The policy choice of Property acquisition seems to perform worse in comparison to the two alternatives and the no-policy baseline across most of the indicators \cite{alves_furtado2022}.

PS2, however, is developed and evaluated for the case of Brasília Metropolitan Region alone. Although the authors also run the model and present comparisons to four other MRs\footnote{Belo Horizonte, Campinas, Fortaleza and Porto Alegre.}, results were not exhaustively run for all of the 46 available MRs. This is exactly how the surrogate model we present contribute to the analysis. We use the bulk of runs of the sensitivity analysis to expand the results to all 46 MRs, while also expanding the variability contained in the data--better exploring the parameter space--,and steering results towards optimal policies.

\subsection{Data}
In order to emulate the original PS2 ABM, we started out with two sets of data produced in the 11,076 runs the authors of PS2 executed \cite{alves_furtado2022}. The data provided by the authors included 32 sets of sensitivity analysis on different parameters (typically with 7 interval values and 20 runs each), runs for all MRs, runs for specific MRs, and sets of 20 runs for each one of the four alternative policy instrument.\footnote{We estimate that all 11,076 runs, on simultaneous 20 cores server would take nearly 300 hours to complete.} For all runs, we accessed the parameters' configuration for each one, as well as the the outputs--in terms of indicators produced by the model--associated with each configuration of parameters (see the left ABM portion of figure \ref{fig:surrogate_ABM_ML}). 

The configuration of parameters and rules includes information such as which Metropolitan Region is being simulated; the percentage of the total population used for a given run\footnote{The baseline model uses only a fraction of the total population to run the default configuration.}; the choice of wage-decision the firms used; or which policy is tested, if any. The full list of parameters include: (a) all of those in table \ref{tab:params}, (b) the Boolean choices of table \ref{tab:S-O-N_dummies}, and (c) each MR, as listed in table \ref{tab:S-O-N_acps}. Moreover, parameters include a test of tax distribution, interest rate alternatives, availability of construction lots, and simulation starting year.

Additionally, PS2 produced an array of 66 indicators, providing details that refer to municipalities of a MR, but that also contained information on households, workers, businesses, and the bank. All in all, our data mapped the exact configuration of parameters for each run to the set of indicators produced by that same run. We then proceeded to the learning process that maps inputs to outputs and enables extrapolation using the surrogate ML model. 

\subsubsection{Concept and design of the optimal}
Here comes the "towards policy" part of the paper. Instead of just learning patterns from data (unsupervised machine learning), we have opted to apply a supervised ML to construct a surrogate model that purposefully steers towards policy. A model that comparatively distinguishes whether the tested policies--and their associated parameters--are socially beneficial or not.

A large portion of public economic debate surpasses the growth of the economy, and adopts GDP as its main indicator. There has been some attempts to qualify the discussion and include other dimensions, such as a Human Development Index (HDI), a focus on the United Nations Sustainable Development Goals (UN-SDG) \cite{guerrero_policy_2020,guerrero_how_2022}, or even include happiness as a subjective indicator, although with limited reach \cite{austin_well-being_2016}.

We decided to focus simultaneously on GDP (production) and Gini coefficient (inequality) on the surrogate model. The algorithm enables the  surrogate with any choice of pair of indicators available in the original model, along with their threshold quantiles. The other output indicators available included: inflation, unemployment, household consumption and wealth, infrastructure investment, GDP and Gini coefficient. Given the alternatives, we considered that the pair of indicators that conveys the stronger combination of optimal policy were the greatest GDP with the smallest Gini coefficient. As such, the economy shows strength at the same time that distributes its growth.\footnote{Consider, for example, unemployment and inflation. Although relevant indicators, by themselves--we assume--they do not represent socially optimal. We might have a stagnant economy with low inflation and unemployment, but with negative growth and high inequality. \cite{alves_furtado2022} defines the Gini coefficient based on permanent income. That includes both previous average household income and value of housing properties or mortgages.} 
2022:39:
Hence, the binary classification of optimal and non-optimal encompasses all simulation runs--out of the total sample of 11,076--that simultaneously qualifies along the \textbf{superior quartile} (.75) of GDP indicator and the \textbf{inferior quartile} (.25) of the Gini coefficient indicator. The quartiles were chosen to reflect a reasonable number of runs as optimal.

The optimal in this case is designed as a comparative and relational statistic among the set of 46 MRs. As such, we may say that optimal refers to a national bundle that includes larger and smaller MR with diverse historical attributes, economic prevalence, population composition, businesses development and accessibility. 

The idea is that the surrogate model enables a counterfactual optimum as an would-be \textit{ex-post} analysis, before actually implementing any of the policies. Given a photo of the existing attributes of each Metropolitan Region, the surrogate responds which policy--if any--would change the \textit{status quo} for the better, given the national context.

\subsubsection{Generation of parameters}\label{generationparameters}
Parameters are generated according to two processes:

\begin{itemize}
    \item If they are discrete parameters or rules, they are chosen according to a probability that is simply the inverse of the number of alternatives ($m^{-1}$), being $m$ the number of possibilities for each dummy analyzed. In the case of policies, for example, each one of the three policies tested, along with the no-policy baseline, had a 0.25\% probability.
    \item If they are continuous parameters, they are drawn from a truncated normal distribution, given the lower and upper bounds (see table \ref{tab:max_min_params}), the mean, and a three times larger standard deviation from the original sample.
\end{itemize}

\subsection{Procedures}\label{sub:procedures}
\begin{figure}[!t]
    \centering
    \includegraphics[width=\linewidth]{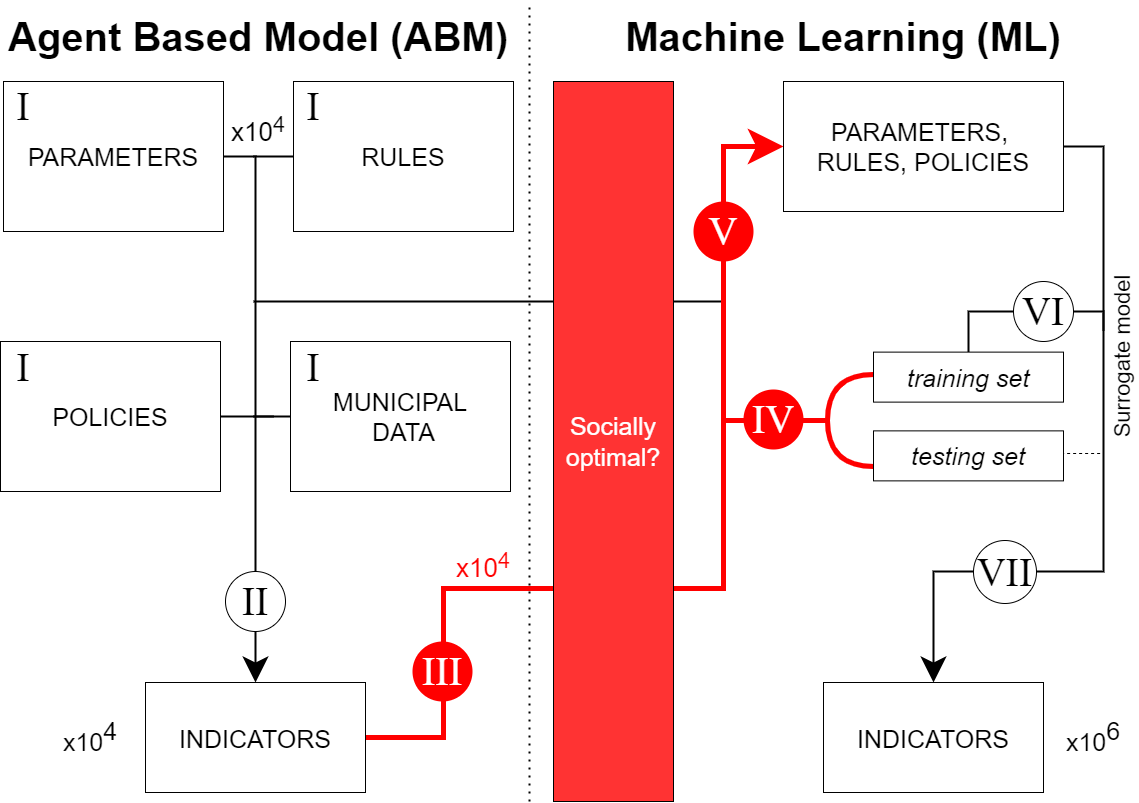}
    \caption{The figure depicts the steps taken to emulate the original ABM, whilst amplifying the possible combination of parameters and steering the results towards optimal policy. The original ABM (left side) produced sample configuration parameters and output indicators (I and II). Indicators were then divided between optimal and non-optimal depending on their relative GDP and Gini coefficient indicators (III). A set of 1,000,000 new parameters were generated and trained in the ML surrogate model, producing one million results (IV--VIII). }
    \label{fig:surrogate_ABM_ML}
\end{figure}

The step-by-step of the implementation of the surrogate is as follows (figure \ref{fig:surrogate_ABM_ML}). The code for the steps listed below is available at  
\url{https://www.comses.net/codebase-release/a336c854-d89d-44a1-adbc-50ae092e64de/}. Please refer to the README.md file for instructions on how to run the code: 

\begin{enumerate}[label=\Roman*.]
    \item The PS2 ABM was manually run by the original authors, producing 11,076 individual runs. Each run contains the configuration parameters: (a) including exogenous parameters, (b) rules and decision-making choices, (c) each one of tested policies -- or the no-policy baseline, (d) the necessary initial municipal input data, and the output of the model containing all final indicators of the simulation. Examples  of each one include:
        \begin{itemize}
            \item Exogenous parameters: workers' productivity, mortgage interests, businesses' markup, or the relevance of neighborhood quality perception in housing prices.
            \item Rules and decision-making mechanisms: whether to consider distance to the firm as a job application criteria, how to set workers' wage, or how to distribute metropolitan funds.
            \item Policies: each one of the three tested policy, plus the no-policy baseline.
            \item Municipal data: local information on households composition, workers qualification, businesses, and the intraurban spatial configuration. Municipal data is constant. However, one of the parameters determines the initial year, and thus the original data, to be either 2010 (default), or 2000.
        \end{itemize}
    \item We then mapped the input configuration parameters to the output indicators, constructing the typical ML X matrix. Hence, every configuration of parameters is linked to the output of the simulation depicted by the indicators.
    \item Next, we constructed the optimal and non-optimal from output indicators, thus producing the typical ML Y vector.
    \item We proceeded to the separation of original sample between training and test sets using the typical python scikit-learn (skelarn) model selection procedure of train-test split, with test size of .25 and random state of 10. Then, we trained the model and tested it (see table \ref{tab:confusion}).
    \item Afterwards, we generated 1,000,000 new parameters, expanding the original configuration parameters into one million combinations (see subsection Methods: Generation of parameters).
    \item We applied the learning algorithm. Running of the ML random forest itself (see subsection Methods: ML implementation).
    \item Then, we applied the learned ML algorithm to the newly generated parameters to find optimal or non-optimal results. 
    \item Finally, we analyzed the results.
\end{enumerate}

In a nutshell, the ABM and the ML surrogate model are separated in their inner-workings. First, the ABM itself runs and produces outputs from configuration parameters, with some variation. Secondly, the output indicators enable the optimal classification. Finally, a much larger configuration of parameters uses the learned ML model--the surrogate--to generate new outputs.

We implemented and tested some ML implementations, such as support-vector machine (SVC), Multi-layer Perceptron Classifier (MLPC) and a Voting Classifier. The SVC and MLPC were unable to predict optimal cases. See from table \ref{tab:confusion2} in the appendix that their optimal predictions were observed to be in fact non-optimal. These two unadjusted results influenced the Voting Classifier, which is a weighted score of the models tested.

Hence, we chose the Random Forest Classifier method that was able to correctly predict optimal and non-optimal cases. We applied the Random Forest using $10^4$ trees, Gini criterion, with a maximum depth of 15 for each tree. The confusion matrix is presented at table \ref{tab:confusion}. The results obtained led to an accuracy of 0.9917 with a precision of 0.9863. However, the recall was not so good at 0.8727, with an F1 indicator of 0.9260. Considering that these results were obtained after 11,076 time-consuming simulation runs, we understand that the surrogate is adequate for the exercise proposed.

The simulation run of the baseline model takes about 30 minutes in the standard server of our institution. The default configuration also allows for simultaneous runs on different cores, which enables 20 runs in the same 30 minutes. Even if we had 100 cores and ran 100 simulations at the same time, we would still need 300,000 minutes, or 5,000 hours, which represents more than 208 days. The surrogate equivalent runs in less than 2--3 hours.

We tried to avoid errors and artefacts in the sense proposed by \cite{galan_errors_2009}. \cite{alves_furtado2022} claim to have made extensive sensitivity analysis of parameters and rules, along with multiple simulations, and also makes the code and documentation open. Our results reinforce the robustness of the ABM output, with a wider parameter space. The application of the ML surrogate is straightforward and uses typical python/scikit-learn implementation. Nevertheless, we provide the code and data. Please refer to the README.md file for instructions on how to run the code: 
\url{https://www.comses.net/codebase-release/a336c854-d89d-44a1-adbc-50ae092e64de/}. 

\begin{table}[!h]
\centering
\caption{Confusion matrix for the random forest ML implementation. The random forest was the model with best accuracy--0.9917--predicting only 2 optimal cases which were actually non-optimal, whereas 21 cases that were actually optimal were predicted to be non-optimal.}
\begin{tabular}{lllr}
\cline{1-4}
                          &             & \multicolumn{2}{l}{Predicted}                          \\ \cline{3-4} 
                          &             & Non-optimal              & \multicolumn{1}{l}{Optimal} \\ 
\multirow{2}{*}{Observed} & Non-optimal & \multicolumn{1}{r}{2602} & 2                           \\ \cline{2-4} 
                          & Optimal     & \multicolumn{1}{r}{21}   & 144                         \\ \hline
\end{tabular}
\label{tab:confusion}
\end{table}
 
 \section{Results and Discussion}\label{results}
In this section, we first present the results of the cities classification across each of the competing policies. Then, we detail the difference in parameters and rules given by the optimal and non-optimal classification. 

\subsection{Metropolitan Regions results per policy}
\begin{figure}[!t]
    \centering
    \includegraphics[width=\textwidth]{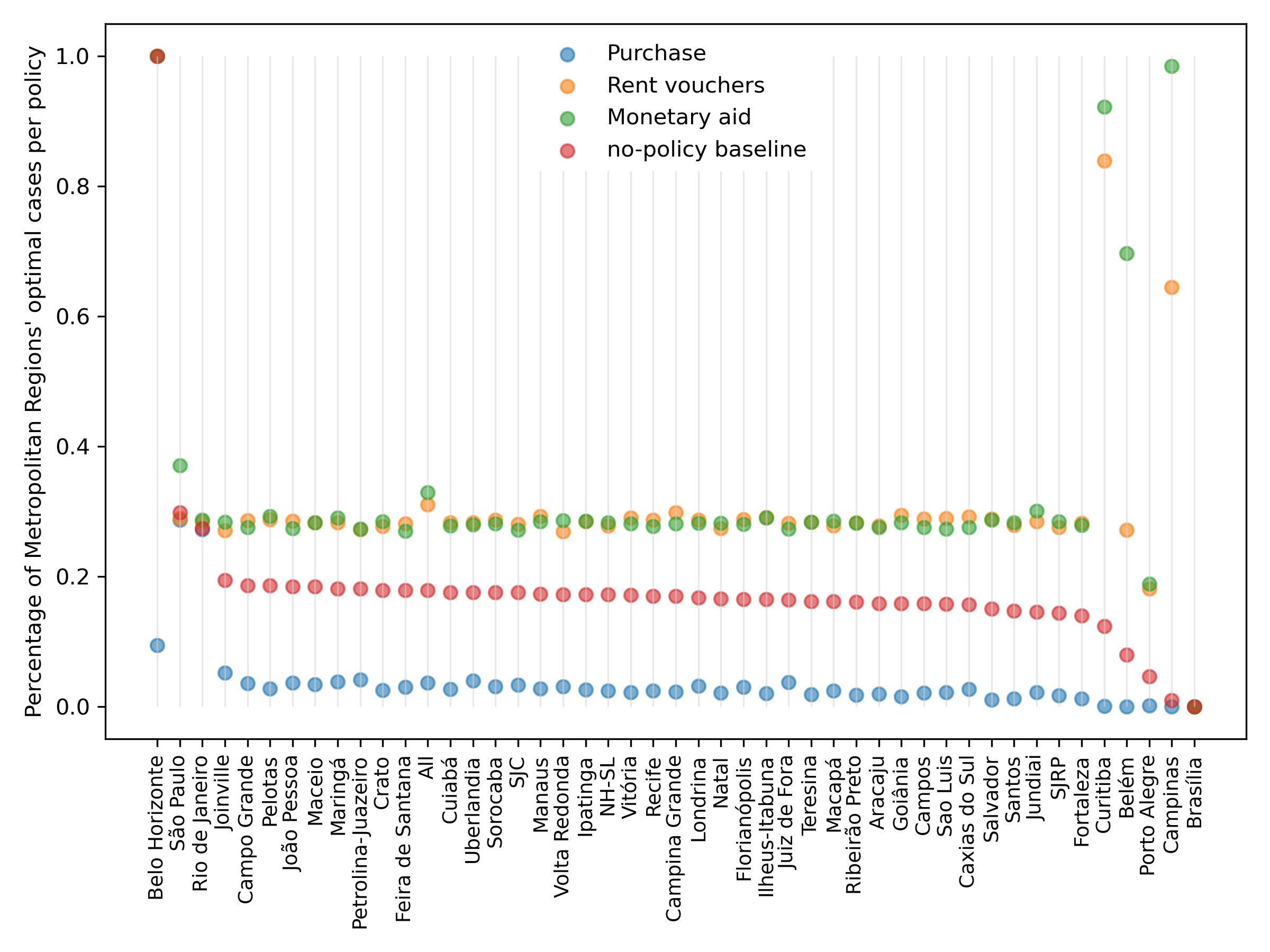}
    \caption{Percentage of optimal cases for the surrogate results, each of the tested policies and the no-policy baseline. The no-policy baseline (in red) clearly separates the mostly negative policy of Property acquisition (blue) and the alternating better policies of Rent vouchers (orange) and Monetary aid (green). }
    \label{fig:sorted_policies}
\end{figure}

Figure \ref{fig:sorted_policies} shows the performance of all MRs for each policy in terms of percentage of optimal runs, given the optimum definition of best quartile of GDP and lowest quartile of Gini coefficient\footnote{All of the MRs rejected the null hypothesis of the Welch’s t-test of equal means between optimal and non-optimal results.}. On average (see table \ref{tab:mean_acp}), 17.87\% of the simulations performed optimally for the no-policy baseline. Three MRs performed relatively better than the others (Belo Horizonte, São Paulo and Rio de Janeiro), whereas four others performed worse (Belém, Porto Alegre, Campinas and Brasília). All of the remaining 39--most with smaller total population (see table \ref{tab:mrs_data})--were classified as optimal between 19\% and 12\% of their own runs. 

Comparatively to the no-policy baseline and in tune with the results found by \cite{alves_furtado2022}, the Property acquisition policy performs worse than the other alternatives for all cases, although with varying intensity. On average, Property acquisition policy implies a loss of 14.25 percentage points relative to the no-policy baseline (table \ref{tab:mean_acp}). Nevertheless, for three cities (Rio de Janeiro, São Paulo and Campinas) the loss is small, around one percentage point, wheres varying between 12 to 15 negative points in most of the other Metropolitan Regions. 

Conversely, both--very different--policy alternatives of Rent vouchers or Monetary aid perform better than the no-policy baseline (figure \ref{fig:sorted_policies}). On average, Monetary aid is slightly better, surpassing the no-policy baseline by 15.05 percentage points. Rent vouchers in turn provide a gain of 13.16 p.p. Across the MRs, however, the performance varies with Rent vouchers being the best option for 24 MRs, and Monetary aid for other 19 MRs, with one tie. 

There are some special cases. Given the design and conception of the optimal and their comparative nature across MRs, two cases are extreme. Brasília--the federal capital and host to a number of well-paid civil servants--has the highest structural inequality (Gini coefficient)--of all MRs. Hence, Brasília does not have any classification as optimal in the surrogate, although it accounts for 2.19\% of the 1,000,000 runs (see table \ref{tab:S-O-N_acps}). In the ABM runs, which Brasília accounts for 48.67\% of the total, it does reach an optimal case in 1.06\% of the time. Our interpretation is not that policymakers should refrain from applying policy in Brasília. Quite on the contrary, applying the policies it would have seem to be not sufficient, given the structural \textit{status quo} relative to all other MRs.  

Conversely, Belo Horizonte seems to benefit from a relatively large GDP, without a largely unequal population (Gini coefficient), thus it reaches the optimal cases in all of the no-policy, Rent vouchers and Monetary aid surrogate runs. Specifically for the case of Belo Horizonte, the Property acquisition policy is especially harmful, reducing the optimal in 90.6 percentage points. 

Curitiba, Belém and Campinas are MRs located close to the optimal in such a way that Rent vouchers and Monetary aid would make a huge positive difference within the national comparison. Curitiba would gain nearly 80 percentage points with Monetary aid policy and 71.6 with Rent vouchers. Belém would improve 61.7 with Monetary aid and Campinas would gain more than 95.7 points also with the Monetary aid policy(see table \ref{tab:mean_acp}).

Table \ref{tab:std_acp} shows the standard-deviation of all MRs, and the average (All). The low level of optimals for the Property acquisition policy is more certain (smaller, more restricted deviation) that for both Rent vouchers and Monetary aid policies. The chance to obtain non-optimal results with the consistently worse policy option (Property acquisition) is higher than the chance to obtain optimal results with the other two options (Monetary aid and Rent vouchers). Hence, it is more certain to fail with Property acquisition than to succeed with either of the other policy instruments. 

Property acquisition policy, for most MRs (with the exception of São Paulo and Rio de Janeiro), almost always leads to non-optimal results. Nevertheless, a Property acquisition or Monetary aid policy is not as certain due to the higher standard deviations. This reinforces the understanding that other factors are relevant to determine the outcome of a policy, rather than 
the policy instrument itself. However, policymakers can be assured that the Property acquisition policy remains as a distant third-best option, unsuited for most MRs and consistently worse than even the no-policy baseline scenario.

\subsection{Parameters and rules analysis}
The authors of the original ABM model calibrated and validated it towards a reasonable performance along four macroeconomic indicators and the spatial distribution of the real estate market for the city of Brasília in the no-policy baseline scenario \cite{alves_furtado2022}. We used the ABM runs' results to calculate which ones fall into our optimal category, as an \textit{ex-post} indicator. 

The comparative nature of the design differentiates between optimality and non-optimality among the MRs. As such, the results represent the average of the influence of the parameters, rules and policies on the ensemble of all MRs. This helps policymakers comprehend--considering larger and smaller, richer and poorer MRs--direction and intensity of better quality policy.

\begin{figure}[!t]
    \centering
    \includegraphics[width=\textwidth]{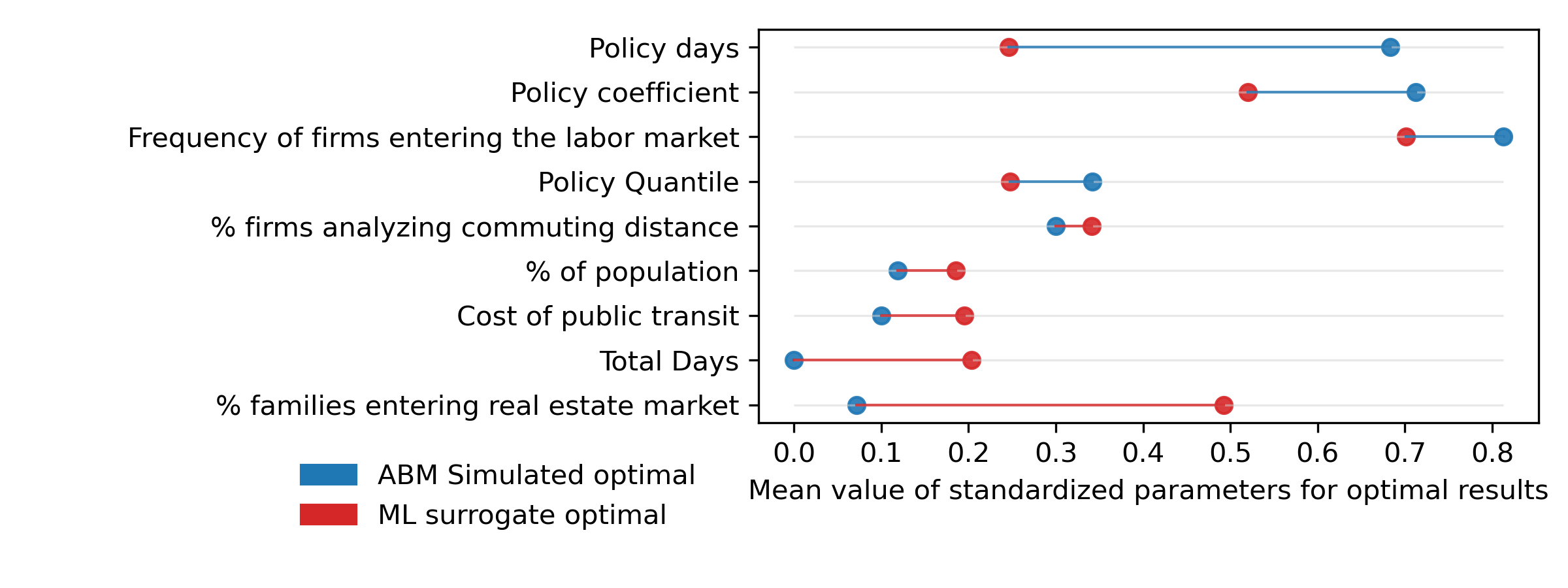}
    \caption{Comparative standard score of parameters for optimal results in relation to the full sample for ABM and ML surrogate runs. We included the parameters for which we could reject the null hypothesis that the means were statistically the same between ABM and ML and also between optimal and non-optimal ML surrogate. The difference from the validated ABM simulation run and the ML surrogate optimal case suggests the counterfactual changes needed in the parameters in order to increase the number of optimal results. The figure is sorted by the difference between ABM Simulated model and ML surrogate optimals. }
    \label{fig:parameters}
\end{figure}

The optimal of our surrogate ML in turn follows the same (learned) mechanisms of the original ABM, but applies instead a varied parameters' space. This procedure enables specific parameter values to be associated with optimal/non-optimal results. That is why we consider which parameters and rules (when possible) would need to be modified were policymakers to nudge their MRs towards better results. We used the standardized value for optimal runs to compare the ABM and ML surrogate. There are higher, similar and smaller values of the standard score for the ABM when compared to the ML surrogate optimal (see figure \ref{fig:parameters}). 

We could also not reject the null hypothesis that the means were the same for some of the parameters. That indicates that either those parameters were already at an optimal level at the ABM calibration, or that they are not much relevant to generate optimal results.

The three parameters that refer to the design of the size of the households' set that are to receive policy aid should have smaller values according to the ML surrogate optimal results. 
Together these parameter optimal values suggest that the choice of families to be included could be more precise and focused. We comment each one in turn. 

Policy days refers to the number of days the municipalities look back in time to check the financial conditions of the households and decide whether to include them in the beneficiary pool. The default value is one year (360 days), but the ABM simulation also tested it for 180 days. ML results suggest then that using a smaller number of backward months in order to include households as policy beneficiaries would generate better overall performance.

Policy coefficient specifies the municipalities' budget share to be invested in the policy. Policy quantile is the income threshold for households to be included in the policy program. Whereas lowering policy coefficient would diminish policy investments, a smaller threshold to include households would probably help focus the policy towards the poorest, most in-need families. 

Conversely, note that the Property acquisition policy was by design a policy that helps a comparatively much smaller number of households, given that the amount of financial resources to Property acquisition houses is considerably higher than that of renting, for example. However, in the case of Property acquisition policy, it seems the number of aided households is too small, which is also not the best solution.

The frequency of firms entering the labor market value is slightly lower in the ML surrogate when compared to the ABM model. This suggests that the optimal is reached more frequently when the labor market is less volatile. 

The parameter of percentage of firms that analyze commute distance as a criteria in the labor market should be slightly higher to enable more optimal results, according to the ML surrogate model results. This parameter refers to the spatial arrangement of the MRs and varies between hiring exclusively via best candidate qualification or via residence proximity to the hiring firm. As discussed in \cite{alves_furtado2022}, it seems that there is a midway arrangement between qualification and commuting distance that best distributes jobs and workers bringing more prosperity for the MR as a whole.

Percentage of population refers to the size of the sample of simulated inhabitants. The higher ML surrogate optimal value depicts the gain that moderately larger MRs have when compared to smaller ones. Total days also indicate that longer simulations may generate more positive feedback effects and help MRs. 

The percentage of households that enter the real estate market is much larger in ML surrogate optimal, compared to that of the ABM surrogate optimal. As expected, a more dynamic real estate market indeed brings economic gains to the MRs. It is relevant to highlight, however, that whereas the ABM considers reasonable inflation, along with other macroeconomic indicators, the ML surrogate is concerned with GDP output and Gini coefficient alone. Yet, the parameter value suggests that dynamic real estate markets may bring positive overall results.

Finally, the ML surrogate optimal suggests that cost of public transit should be a bit larger, when compared to the ABM. As the cost of public transit is considered by the candidate as a criteria when deciding the firm to work with, the results may indicate that a better spatial match between workers and firms--mediated via considering more heavily transport costs--might bring more positive results for the full set of the MRs.

\begin{table}[!t]
\centering
\caption{Comparison of parameters' absolute values for ABM optimal and ML surrogate optimal results. }
\begin{tabular}{lrr}
\toprule
\multirow{2}{*}{Parameters} & \multicolumn{2}{c}{Standardized values}  \\
{}  &  ABM optimal & ML surrogate optimal\\
\midrule
\% firms analyze commute distance             &               0.300 &        0.341 \\
\% of construction firms                      &               0.200 &        0.446 \\
\% of population                              &               0.118 &        0.185 \\
\% that enters the estate market              &               0.071 &        0.492 \\
Cost of private transit                      &               0.500 &        0.498 \\
Cost of public transit                       &               0.100 &        0.195 \\
Frequency of firms entering the labor market &               0.812 &        0.701 \\
Hiring sample size                           &               0.655 &        0.510 \\
Loan/permament income ratio                  &               0.625 &        0.528 \\
Markup                                       &               0.500 &        0.505 \\
Maximum Loan-to-Value                        &               0.625 &        0.532 \\
Municipal efficiency management              &               0.474 &        0.444 \\
Neighborhood effect                          &               0.571 &        0.579 \\
Perceived market size                        &               0.474 &        0.476 \\
Policy Quantile                              &               0.342 &        0.248 \\
Policy coefficient                           &               0.712 &        0.520 \\
Policy days                                  &               0.683 &        0.246 \\
Productivity: divisor                        &               0.314 &        0.566 \\
Productivity: exponent                       &               0.667 &        0.579 \\
Sticky Prices                                &               0.500 &        0.489 \\
Supply-demand effect on real estate prices   &               0.500 &        0.483 \\
Tax over estate transactions                 &               0.667 &        0.514 \\
Total Days                                   &               0.000 &        0.204 \\
\bottomrule
\end{tabular}
\label{tab:params}
\end{table}

\begin{table}[!t]
\centering
\caption{Differences between surrogate and agent-based models in relation to sample size (in \% of total simulations), and both optimal and non-optimal cases (\% of cases out of the total that fall under each category) for selected dummy parameters of the model. }
\begin{tabular}{lrrrrrr}
\hline
\multirow{2}{*}{\begin{tabular}[c]{@{}l@{}}Rules and\\Policies' choices \end{tabular}} & \multicolumn{2}{c}{Size (\%)} & \multicolumn{2}{c}{Optimal (\%)} & \multicolumn{2}{c}{Non-optimal (\%)} \\
 & Surrogate & ABM & Surrogate & ABM & Surrogate & ABM \\ \hline
     Policy: Property acquisition &      25.06 &      12.19 &      3.62 &      2.59 &         96.38 &         97.41 \\
     No-policy baseline &      24.93 &      63.43 &     17.87 &      3.91 &         82.13 &         96.09 \\
     Policy: Rent vouchers &      25.01 &      12.19 &     31.03 &     10.37 &         68.97 &         89.63 \\
     Policy: Monetary aid &      25.00 &      12.19 &     32.92 &     13.93 &         67.08 &         86.07 \\ \hline
\end{tabular}

\label{tab:S-O-N_dummies}
\end{table}

\subsection{Discussion}
In practical terms, the exercise proposed generalizes earlier results and pinpoints which policy is best for which city. Moreover, the robustness of the results confirm that the Property acquisition policy performs worse in terms of GDP and Gini coefficient within a varied input of MRs. Rent vouchers and Monetary aid provide similar results, but may be different for specific MRs. 

The federal government (and policymakers) may learn from the analysis performed. The results provide quantitative and empirical data that support alternative emphasis on policy instruments. Typically, Brazilians and OECD members prefer to own their houses \cite{causa_housing_2019}. Indeed the Property acquisition program in Brazil was called My House, My Life (PMCMV).\footnote{Originally, in Portuguese, Minha Casa Minha Vida.} However, the program has been criticized by the lack of city integration (distant locations and without infrastructure) \cite{santo_amore_minha_2015}, and has virtually stopped due to difficulties in funding. Nationally, no other housing policy has replaced PMCMV. A few studies have recommended rent vouchers as alternatives \cite{dias_um_2021,rogar_o_2018}. Monetary aid, in turn, has steadily gained support in Brazil \cite{fonseca_bolsa_2018}. Our robust results support the novel comparison among housing and social welfare policy instruments making it explicit for policymakers how they compare and what benefits are reached, considering GDP and Gini coefficient as policy goals. 

The findings of the surrogate contributes to policy analysis literature by showing that other alternatives (to the Property acquisition policy) not only are available, but also might be more beneficial. PMCMV--the large Property acquisition program in Brazil--was decided upon without observing the planning legislation that was in place, nor the opinion of urban policy experts \cite{ferreira_politica_2019}. 

The findings of the comparison among the ABM and the ML surrogate optimal results make sense and quantitatively inform policymakers about the direction and intensity of necessary adjustments to replicate optimal MRs performance. All in all, the comparisons suggest: (a) a more focused set of low-income definition of households to assist, (b) with a smaller number of months (of income observation) to include the households, (c) a less volatile (safer) labor market, (d) with a consideration of transport costs to improve the quality of spatial match between firms and employees, (e) and that larger metropolises fair better with increased gains from agglomeration.

\section{Concluding remarks}
We present a Machine Learning (ML) surrogate model that departs from an ABM model and generalizes its results for a wider number of Metropolitan Regions and policy tests. In the process, we include a much larger, more varied sample of possible input parameters and rules. We also steer the ML surrogate model towards a policy optimum that combines larger GDP output and smaller Gini coefficient within the set of all 46 modeled MRs. Besides the actual findings of best-policy for each MR, the exercise also serves as a test of robustness of the original results. 

Considering the research questions, we find that, indeed, different MRs do best in different tested policies. Although on average, the Monetary aid policy performs better, the Rent vouchers is the best policy for a higher number of MRs (24). We further confirm previous results with worst performance for the Property acquisition policy, vis-à-vis those of the no-policy baseline. 

Results also showed that some parameters have statistically significant different optimal results in the ML surrogate model and the ABM model. We interpret these values as those the parameters would have to be transformed into, were the MRs to aspire to more socially optimal positions. 

The paper presents an exercise to expand the combination of parameters, test the robustness of the results, and evaluate possible pathways for policymakers in terms of which parameters would need changing, were they interested in reaching higher GDP with lower Gini coefficient. However, results are limited by the available data used to produce the surrogate model and the intrinsic validated of the original model itself. Conceptually, we show that a reasonably fast policy evaluation of ABM results--"ex post"--is possible, and one that adds information to policymakers.

In terms of future research, we are interested to learn whether intrinsic Metropolitan Region' attributes and structural characteristics translates into a higher probability of getting comparatively optimal results. That is, are household attributes and location, number of firms, population qualification and age composition, municipalities boundaries and their geographical interconnections sufficient to determine relative position of one MR in relation to the others? In order to answer such question, we would need to search for the determinants of the optimum, probably via regression techniques. 

Moreover, we envision a study that uses dimensionality reduction techniques to summarize a larger number of output indicators to better reflect policymakers preferences. Along with a more comprehensive output indicator, future work should take advantage of the possibilities of mixing simultaneous policy instruments, and those from different domains, such as housing policy and social welfare.


\bibliography{labib} 

\section{Appendices}

\subsection{Results of other Machine Learning techniques}

Besides random forest, we also applied a Multi-layer Perceptron Classifier (MLPC) that optimizes the log-loss function \cite{glorot_understanding_2010}, and a C-Support Vector Classification, which relies on a library for Support Vector Machine (libSVM) \cite{chang_libsvm_2011}. Finally, we also used a Voting Classifier implementation that uses other model estimators and applies a majority rule voting. See table \ref{tab:confusion2} for confustion matrix results. 

\begin{table}[!t]
\centering
\caption{Confusion matrix for other ML implementation: Multi-layer Perceptron Classifier (MLPC), C-Support Vector Classification (C-SVC), and Voting Classifier.}
\begin{tabular}{lllr}
\cline{1-4}
                          &             & \multicolumn{2}{l}{Predicted}                          \\  
                          &             & Non-optimal              & \multicolumn{1}{l}{Optimal} \\ \hline
\multirow{2}{*}{Observed MLPC} & Non-optimal & \multicolumn{1}{r}{2604} & 0                           \\ 
                          & Optimal     & \multicolumn{1}{r}{165}   & 0                         \\ \
\multirow{2}{*}{Observed C-SVC} & Non-optimal & \multicolumn{1}{r}{2604} & 0                           \\ 
                          & Optimal     & \multicolumn{1}{r}{165}   & 0                         \\ 
\multirow{2}{*}{Observed Voting} & Non-optimal & \multicolumn{1}{r}{2604} & 0                           \\ 
                          & Optimal     & \multicolumn{1}{r}{134}   & 31                         \\ \hline
\end{tabular}
\label{tab:confusion2}
\end{table}

\subsection{Percentage by policy}
See table \ref{tab:mean_acp} for the percentage points difference of policies relative to the no-policy baseline.

\begin{table}[!t]
\centering
\caption{Mean performance of each policy relative to the \textbf{no-policy baseline} for Brazilian MRs. Results for Property acquisition, Rent vouchers and Monetary aid denote performance superior (positive) and inferior (negative) in terms of percentage points, compared to that same MR for the no-policy baseline. All MRs perform worse when policy Property acquisition is applied. Conversely, Monetary aid result in positive performance for all MRs, with varying levels of improvement.} 
\begin{tabular}{lrrrr} 
\hline
\multirow{2}{*}{Metropolitan Region \textbackslash{} Policy} & \multicolumn{1}{l}{\begin{tabular}[c]{@{}l@{}}\% \\ \end{tabular}} & \multicolumn{3}{l}{\begin{tabular}[c]{@{}l@{}}Difference to the 'no-policy'\\scenario (in percentage points) \end{tabular}} \\ \cline{2-5}
 & \multicolumn{1}{l}{\begin{tabular}[c]{@{}l@{}}No-policy \\(baseline)\end{tabular}} & \multicolumn{1}{l}{Property acquisition} & \multicolumn{1}{l}{\begin{tabular}[c]{@{}l@{}}Rent\\vouchers \end{tabular}} & \multicolumn{1}{l}{\begin{tabular}[c]{@{}l@{}}Monetary\\aid \end{tabular}} \\ 
\cline{1-1}\cmidrule{2-5}
     Belo Horizonte &        100.00 &       -90.58 &          0.00 &          0.00 \\
         São Paulo &         29.84 &        -1.14 &         -0.90 &          7.23 \\
    Rio de Janeiro &         27.38 &        -0.12 &          1.11 &          1.33 \\
         Joinville &         19.43 &       -14.26 &          7.67 &         8.91 \\
      Campo Grande &         18.63 &       -15.02 &         10.02 &          8.91 \\
           Pelotas &         18.58 &       -15.81 &         10.24 &         10.71 \\
       João Pessoa &         18.47 &       -14.78 &         10.11 &          8.97 \\
            Maceió &         18.43 &       -15.06 &          9.84 &          9.83 \\
           Maringá &         18.14 &       -14.33 &         10.16 &         10.86 \\
Petrolina-Juazeiro &         18.11 &       -13.97 &          9.16 &         9.25 \\
             Crato &         17.92 &       -15.45 &          9.81 &         10.55 \\
  Feira de Santana &         17.88 &       -14.90 &         10.29 &          9.08 \\
  \hline
All &         17.87 &       -14.25 &         13.16 &         15.05 \\
\hline
            Cuiabá &         17.59 &       -14.91 &         10.70 &         10.21 \\
        Uberlandia &         17.57 &       -13.62 &         10.73 &         10.39 \\
          Sorocaba &         17.56 &       -14.45 &         11.14 &         10.60 \\
               SJC &         17.54 &       -14.21 &         10.55 &          9.59 \\
            Manaus &         17.30 &       -14.58 &         11.94 &         11.14 \\
     Volta Redonda &         17.23 &       -14.12 &          9.72 &         11.38 \\
          Ipatinga &         17.22 &       -14.61 &         11.26 &         11.34 \\
             NH-SL &         17.21 &       -14.78 &         10.61 &         11.06 \\
           Vitória &         17.18 &       -14.97 &         11.86 &         10.99 \\
            Recife &         17.02 &       -14.62 &         11.65 &         10.69 \\
    Campina Grande &         16.96 &       -14.73 &         12.92 &         11.15 \\
          Londrina &         16.77 &       -13.63 &         11.96 &         11.48 \\
             Natal &         16.62 &       -14.53 &         10.81 &         11.59 \\
     Florianópolis &         16.50 &       -13.53 &         12.25 &         11.57 \\
    Ilheus-Itabuna &         16.50 &       -14.49 &         12.57 &         12.52 \\
      Juiz de Fora &         16.40 &       -12.68 &         11.81 &         10.90 \\
          Teresina &         16.21 &       -14.34 &         12.15 &         12.15 \\
            Macapá &         16.15 &       -13.75 &         11.67 &         12.36 \\
    Ribeirão Preto &         16.08 &       -14.28 &         12.20 &         12.16 \\
           Aracaju &         15.87 &       -13.97 &         11.93 &         11.73 \\
           Goiânia &         15.86 &       -14.32 &         13.58 &         12.45 \\
            Campos &         15.83 &       -13.76 &         13.06 &         11.73 \\
          Sao Luis &         15.73 &       -13.54 &         13.22 &         11.61 \\
     Caxias do Sul &         15.71 &       -13.06 &         13.46 &         11.89 \\
          Salvador &         15.04 &       -14.01 &         13.83 &         13.62 \\
            Santos &         14.67 &       -13.48 &         13.20 &         13.62 \\
           Jundiai &         14.56 &       -12.39 &         13.91 &         15.50 \\
              SJRP &         14.36 &       -12.68 &         13.23 &         14.07 \\
         Fortaleza &         13.99 &       -12.79 &         14.26 &         13.88 \\
          Curitiba &         12.37 &       -12.30 &         71.55 &         79.85 \\
             Belém &          8.00 &        -8.00 &         19.12 &         61.69 \\
      Porto Alegre &          4.62 &        -4.47 &         13.50 &         14.24 \\
          Campinas &          0.94 &        -0.94 &         63.56 &         97.53 \\
          Brasília &          0.00 &         0.00 &          0.00 &          0.00 \\

\hline
\multicolumn{5}{l}{NH-SL: Novo Hamburgo/Sao Leopoldo. SJRP: Sao Jose do Rio Preto.} \\
\multicolumn{5}{l}{SJC: Sao Jose dos Campos}
\end{tabular}

\label{tab:mean_acp}
\end{table}

\subsection{Size, optimal and non-optimal percentages}
See table \ref{tab:S-O-N_acps} for the sizes of samples, optimal and non-optimal results. 

\begin{table}[!t]
\centering
\caption{Results from surrogate and agent-based models in relation to sample size (in \% of total simulations), and both optimal and non-optimal cases (\% of cases out of the total that fall under each category) for MRs of the model. }
\begin{tabular}{l|rrrrrr}
\hline
\multirow{2}{*}{Metropolitan Region} & \multicolumn{2}{c}{Size (\%)} & \multicolumn{2}{c}{Optimal (\%)} & \multicolumn{2}{c}{Non-optimal (\%)} \\
 & Surrogate & ABM & Surrogate & ABM & Surrogate & ABM \\ \hline
Aracaju &       2.21 &       1.06 &     18.16 &      0.00 &         81.84 &        100.00 \\
    Belo Horizonte &       2.17 &       2.00 &     77.06 &     81.08 &         22.94 &         18.92 \\
             Belém &       2.18 &       1.06 &     26.49 &     21.37 &         73.51 &         78.63 \\
          Brasília &       2.19 &      48.67 &      0.00 &      1.06 &        100.00 &         98.94 \\
    Campina Grande &       2.16 &       1.06 &     19.29 &      0.00 &         80.71 &        100.00 \\
          Campinas &       2.18 &       2.00 &     41.01 &     32.88 &         58.99 &         67.12 \\
      Campo Grande &       2.15 &       1.06 &     19.54 &      0.00 &         80.46 &        100.00 \\
            Campos &       2.17 &       1.06 &     18.41 &      0.00 &         81.59 &        100.00 \\
     Caxias do Sul &       2.18 &       1.06 &     18.94 &      0.00 &         81.06 &        100.00 \\
             Crato &       2.18 &       1.06 &     19.07 &      0.00 &         80.93 &        100.00 \\
            Cuiabá &       2.16 &       1.06 &     19.10 &      0.00 &         80.90 &        100.00 \\
          Curitiba &       2.15 &       1.06 &     46.99 &     47.86 &         53.01 &         52.14 \\
  Feira de Santana &       2.16 &       1.06 &     18.89 &      0.00 &         81.11 &        100.00 \\
     Florianópolis &       2.18 &       1.06 &     19.18 &      0.00 &         80.82 &        100.00 \\
         Fortaleza &       2.17 &       2.00 &     17.95 &      3.15 &         82.05 &         96.85 \\
           Goiânia &       2.16 &       1.06 &     18.76 &      0.00 &         81.24 &        100.00 \\
    Ilheus-Itabuna &       2.18 &       1.06 &     19.20 &      0.00 &         80.80 &        100.00 \\
          Ipatinga &       2.17 &       1.06 &     19.25 &      0.00 &         80.75 &        100.00 \\
         Joinville &       2.16 &       0.30 &     19.99 &      0.00 &         80.01 &        100.00 \\
       João Pessoa &       2.18 &       1.06 &     19.52 &      0.00 &         80.48 &        100.00 \\
      Juiz de Fora &       2.16 &       1.06 &     18.78 &      0.00 &         81.22 &        100.00 \\
           Jundiai &       2.18 &       1.06 &     18.80 &      0.00 &         81.20 &        100.00 \\
          Londrina &       2.18 &       1.06 &     19.24 &      0.00 &         80.76 &        100.00 \\
            Macapá &       2.18 &       1.06 &     18.82 &      0.00 &         81.18 &        100.00 \\
            Maceio &       2.16 &       1.06 &     19.66 &      0.00 &         80.34 &        100.00 \\
            Manaus &       2.17 &       1.06 &     19.42 &      0.00 &         80.58 &        100.00 \\
           Maringá &       2.16 &       1.06 &     19.90 &      0.00 &         80.10 &        100.00 \\
             NH-SL &       2.16 &       1.06 &     18.98 &      0.00 &         81.02 &        100.00 \\
             Natal &       2.19 &       1.06 &     18.54 &      0.00 &         81.46 &        100.00 \\
           Pelotas &       2.16 &       1.06 &     19.81 &      0.00 &         80.19 &        100.00 \\
Petrolina-Juazeiro &       2.18 &       1.06 &     19.21 &      0.00 &         80.79 &        100.00 \\
      Porto Alegre &       2.20 &       2.00 &     10.52 &      0.00 &         89.48 &        100.00 \\
            Recife &       2.17 &       1.06 &     18.92 &      0.00 &         81.08 &        100.00 \\
    Ribeirão Preto &       2.16 &       1.06 &     18.73 &      0.00 &         81.27 &        100.00 \\
    Rio de Janeiro &       2.19 &       1.81 &     27.95 &     54.73 &         72.05 &         45.27 \\
               SJC &       2.18 &       1.06 &     18.95 &      0.00 &         81.05 &        100.00 \\
              SJRP &       2.17 &       1.06 &     17.97 &      0.00 &         82.03 &        100.00 \\
          Salvador &       2.17 &       1.06 &     18.27 &     11.11 &         81.73 &         88.89 \\
            Santos &       2.17 &       1.06 &     18.01 &      0.00 &         81.99 &        100.00 \\
          Sao Luis &       2.17 &       1.06 &     18.53 &      0.00 &         81.47 &        100.00 \\
          Sorocaba &       2.17 &       1.06 &     19.26 &      0.00 &         80.74 &        100.00 \\
         São Paulo &       2.19 &       1.06 &     31.10 &    100.00 &         68.90 &          0.00 \\
          Teresina &       2.19 &       1.06 &     18.81 &      0.00 &         81.19 &        100.00 \\
        Uberlandia &       2.18 &       1.06 &     19.43 &      0.00 &         80.57 &        100.00 \\
           Vitória &       2.20 &       1.06 &     19.20 &      0.00 &         80.80 &        100.00 \\
     Volta Redonda &       2.16 &       1.06 &     19.07 &      0.00 &         80.93 &        100.00 \\ \hline 
     \multicolumn{7}{l}{NH-SL: Novo Hamburgo/Sao Leopoldo} \\
\multicolumn{7}{l}{SJRP: Sao Jose do Rio Preto} \\
\multicolumn{7}{l}{SJC: Sao Jose dos Campos}
\end{tabular}

\label{tab:S-O-N_acps}
\end{table}

\subsection{Standard-deviation}
See table \ref{tab:std_acp} for the standard-deviation of the results for each Metropolitan Region.

\subsection{Parameters boundaries configuration}

\begin{table}[!t]
\centering
\caption{Standard deviation of the no-policy baseline compared to policy performance for Brazilian MRs.}
\begin{tabular}{lrrrr}
\hline
\multirow{2}{*}{Metropolitan Region \textbackslash{} Policy} & \multicolumn{1}{l}{\begin{tabular}[c]{@{}l@{}}\% \\ \end{tabular}} & \multicolumn{3}{l}{\begin{tabular}[c]{@{}l@{}}Difference to the 'no-policy'\\scenario (in percentage points) \end{tabular}} \\ 
\cline{2-5}
 & \multicolumn{1}{l}{\begin{tabular}[c]{@{}l@{}}No-policy \\(baseline)\end{tabular}} & \multicolumn{1}{l}{Property acquisition} & \multicolumn{1}{l}{\begin{tabular}[c]{@{}l@{}}Rent\\vouchers \end{tabular}} & \multicolumn{1}{l}{\begin{tabular}[c]{@{}l@{}}Monetary\\aid \end{tabular}} \\ 
\cline{1-1}\cmidrule{2-5}
São Paulo &         45.75 &        -0.51 &         -0.40 &          2.55 \\
    Rio de Janeiro &         44.59 &        -0.06 &          0.55 &          0.65 \\
         Joinville &         39.56 &       -17.41 &          4.89 &          5.51 \\
      Campo Grande &         38.94 &       -20.30 &          6.27 &          5.73 \\
           Pelotas &         38.90 &       -22.48 &          6.39 &          6.61 \\
       João Pessoa &         38.80 &       -19.95 &          6.38 &          5.82 \\
            Maceio &         38.78 &       -20.74 &          6.25 &          6.25 \\
           Maringá &         38.54 &       -19.39 &          6.51 &          6.84 \\
Petrolina-Juazeiro &         38.51 &       -18.59 &          6.02 &          6.07 \\
             Crato &         38.35 &       -22.82 &          6.42 &          6.78 \\
  Feira de Santana &         38.32 &       -21.31 &          6.66 &          6.05 \\ \hline
               All &         38.31 &       -19.63 &          7.95 &          8.68  \\ \hline
            Cuiabá &         38.08 &       -21.93 &          6.96 &          6.72 \\
        Uberlandia &         38.06 &       -18.59 &          6.99 &          6.82 \\
          Sorocaba &         38.05 &       -20.70 &          7.19 &          6.93 \\
               SJC &         38.03 &       -20.09 &          6.91 &          6.43 \\
            Manaus &         37.82 &       -21.56 &          7.67 &          7.29 \\
     Volta Redonda &         37.77 &       -20.41 &          6.60 &          7.42 \\
          Ipatinga &         37.75 &       -21.81 &          7.38 &          7.42 \\
             NH-SL &         37.75 &       -22.36 &          7.06 &          7.28 \\
           Vitória &         37.72 &       -23.02 &          7.68 &          7.26 \\
            Recife &         37.58 &       -22.28 &          7.64 &          7.18 \\
    Campina Grande &         37.53 &       -22.75 &          8.24 &          7.42 \\
          Londrina &         37.36 &       -19.91 &          7.89 &          7.66 \\
             Natal &         37.23 &       -22.91 &          7.39 &          7.77 \\
     Florianópolis &         37.12 &       -20.15 &          8.14 &          7.81 \\
    Ilheus-Itabuna &         37.11 &       -23.08 &          8.30 &          8.28 \\
      Juiz de Fora &         37.03 &       -18.12 &          7.97 &          7.52 \\
          Teresina &         36.85 &       -23.30 &          8.22 &          8.23 \\
            Macapá &         36.80 &       -21.49 &          8.01 &          8.35 \\
    Ribeirão Preto &         36.74 &       -23.44 &          8.29 &          8.28 \\
           Aracaju &         36.54 &       -22.87 &          8.26 &          8.16 \\
           Goiânia &         36.53 &       -24.20 &          9.05 &          8.52 \\
            Campos &         36.50 &       -22.25 &          8.82 &          8.18 \\
          Sao Luis &         36.41 &       -21.76 &          8.94 &          8.16 \\
     Caxias do Sul &         36.39 &       -20.32 &          9.06 &          8.31 \\
          Salvador &         35.75 &       -25.64 &          9.57 &          9.47 \\
            Santos &         35.38 &       -24.54 &          9.45 &          9.66 \\
           Jundiai &         35.27 &       -20.71 &          9.86 &         10.58 \\
              SJRP &         35.07 &       -22.23 &          9.63 &         10.04 \\
         Fortaleza &         34.69 &       -23.80 &         10.33 &         10.15 \\
          Curitiba &         32.92 &       -30.22 &          3.81 &         -6.13 \\
             Belém &         27.12 &       -27.12 &         17.34 &         18.84 \\
      Porto Alegre &         20.99 &       -17.11 &         17.53 &         18.13 \\
          Campinas &          9.63 &        -9.63 &         38.22 &          2.64 \\
          Brasília &          0.00 &         0.00 &          0.00 &          0.00 \\
    Belo Horizonte &          0.00 &        29.21 &          0.00 &          0.00 \\ \hline
\multicolumn{5}{l}{NH-SL: Novo Hamburgo/Sao Leopoldo} \\
\multicolumn{5}{l}{SJRP: Sao Jose do Rio Preto} \\
\multicolumn{5}{l}{SJC: Sao Jose dos Campos}
\end{tabular}

\label{tab:std_acp}
\end{table}

See table \ref{tab:max_min_params} for the boundary restrictions for each parameter of the simulation. Each parameter was restricted to possible values, although considering a wider range of variation. For example, markup is bounded between 0 and a 50\% increase on prices, considering the default model value of 15\%.

We shall briefly explain some parameters, although \cite{alves_furtado2022} provides further explanation. 

Capped low and top values for the bank refer to the total amount of resources available for the bank. 

Policy quantile refers to the size of portion of the poorest families to be included as policy recipients. 

Decay factor refers to the loss of property value with time, following equation \ref{eq: decay}.

\begin{equation}
value = (1- Discount) * e^{Decay*t} + Discount
\label{eq: decay}
\end{equation}

where, $Discount$ refers to the maximum offer discount, $t$ refers to the time that such property has been on the market, and $Decay$ is the decay factor of property value. 

Monthly revenue installments division refers only to construction firms, and it refers to the number of months firms take to incorporate revenue after sales. This is relevant only to determine employees' wages.

Available lots per Neighbourhood, in the model "T\_LICENSES\_PER\_REGION" is a monthly designation of licenses for urbanized lots to become available for construction. 

Unemployment refers to whether firms observe unemployment or not when setting wages. 

Finally, Alternative0 and FPM distribution refer to a previously policy test made by \cite{furtado_policyspace:_2018}.

\begin{table}[!t]
\centering
\caption{Maximum and minimum values or alternatives for simulation parameters. }
\begin{tabular}{l|rr} 
\hline
Parameter & Max & Min \\ 
\hline
Productivity: exponent & 1 & 0 \\
Productivity: divisor & 20 & 1 \\
Municipal efficiency management & 0.001 & 0.00001 \\
Markup & 0.5 & 0 \\
Sticky Prices & 1 & 0 \\
Perceived market size & 100 & 1 \\
Frequency of firms entering the labor market & 1 & 0 \\
\% firms analyzing commuting distance & 1 & 0 \\
Hiring sample size & 100 & 1 \\
Tax: consumption & 0.6 & 0.1 \\
Tax: labor & 0.6 & 0.01 \\
Tax over estate transactions & 0.01 & 0.0001 \\
Tax: firm & 0.6 & 0.01 \\
Tax: property & 0.01 & 0.0001 \\
Policy coefficient & 0.4 & 0 \\
Policy days & 3600 & 0 \\
Policy Quantile & 1 & 0 \\
Age cap for borrower at end of contract & 100 & 50 \\
Loan/permament income ratio & 1 & 0 \\
Maximum Loan-to-Value & 1 & 0 \\
Bank resources: maximum & 1 & 0 \\
Value cap for banks: top & 2 & 1 \\
Value cap for banks: bottom & 1 & 0 \\
Supply-demand effect on real estate prices & 5 & 0 \\
Decay factor for properties & 0 & -0.1 \\
Maximum offer discount & 1 & 0.4 \\
\% families entering real estate market & 0.01 & 0 \\
Neighborhood effect & 5 & 0 \\
Rental Share & 1 & 0 \\
Initial rental price & 0.01 & 0 \\
\% of construction firms & 0.2 & 0 \\
Monthly revenue installments division & 100 & 1 \\
Cost of lots (\% of construction) & 0.7 & 0 \\
Cost of private transit & 0.5 & 0 \\
Cost of public transit & 0.5 & 0 \\
\% of population & 1 & 0 \\
Total Days & 14610 & 1826 \\ 
\hline
Available lots per Neighbourhood & \multicolumn{2}{c}{True, False} \\
Starting day & \multicolumn{2}{c}{2010-01-01, 2000-01-01} \\
Interest & \multicolumn{2}{c}{nominal, real, fixed} \\
Wage is unrelated to unemployment & \multicolumn{2}{c}{True, False} \\
Alternative0 & \multicolumn{2}{c}{True, False} \\
FPM distribution & \multicolumn{2}{c}{True, False} \\
\hline
Policies & \multicolumn{2}{c}{\begin{tabular}[c]{@{}c@{}}No-policy, Property acquisition, \\Monetary aid, Rent vouchers \end{tabular}} \\
\hline
\end{tabular}

\label{tab:max_min_params}
\end{table}

\subsection{MRs Data}
See table \ref{tab:mrs_data} for basic data for each MRs. 

\begin{table}[h]
    \centering
    \caption{Basic data from MRs. Total population for MRs from 2010, Municipal Human Development Index (MHDI), and percentage of population with 13 or more years of study. Obs.: Adapted from input data by \cite{alves_furtado2022} at \url{github.com/bafurtado/policyspace2/tree/master/input}}
    \label{tab:mrs_data}
\begin{tabular}{lrrr}
\toprule
{} &         Total Population 2010 & MHDI & \% College $+$ \\
MRs               &             &        &           \\
\midrule
São Paulo          &  20,096,809 &  0.789 &     0.120 \\
Rio de Janeiro     &  11,968,973 &  0.765 &     0.110 \\
Belo Horizonte     &   5,039,123 &  0.769 &     0.102 \\
Recife             &   3,706,628 &  0.732 &     0.078 \\
Salvador           &   3,503,340 &  0.744 &     0.083 \\
Brasília           &   3,460,637 &  0.793 &     0.110 \\
Fortaleza          &   3,327,021 &  0.733 &     0.073 \\
Curitiba           &   2,993,678 &  0.785 &     0.126 \\
Porto Alegre       &   2,906,148 &  0.767 &     0.120 \\
Campinas           &   2,603,819 &  0.792 &     0.112 \\
Belém              &   2,125,135 &  0.730 &     0.071 \\
Goiânia            &   2,011,735 &  0.768 &     0.106 \\
Manaus             &   1,802,014 &  0.737 &     0.069 \\
Vitória            &   1,670,679 &  0.768 &     0.093 \\
Santos             &   1,604,363 &  0.776 &     0.081 \\
SJC                &   1,392,552 &  0.796 &     0.108 \\
São Luis           &   1,309,330 &  0.754 &     0.072 \\
Natal              &   1,251,459 &  0.740 &     0.083 \\
Sorocaba           &   1,148,035 &  0.783 &     0.096 \\
Maceió             &   1,028,249 &  0.713 &     0.094 \\
João Pessoa        &   1,017,742 &  0.732 &     0.100 \\
Teresina           &     982,968 &  0.733 &     0.085 \\
Aracaju            &     879,061 &  0.731 &     0.087 \\
Florianópolis      &     866,098 &  0.812 &     0.148 \\
Cuiabá             &     803,694 &  0.769 &     0.105 \\
Ribeirão Preto     &     791,295 &  0.788 &     0.137 \\
Campo Grande       &     786,797 &  0.784 &     0.117 \\
Joinville          &     748,470 &  0.799 &     0.107 \\
NH-SL              &     736,405 &  0.739 &     0.057 \\
Londrina           &     709,494 &  0.765 &     0.109 \\
Jundiaí            &     674,877 &  0.795 &     0.106 \\
Uberlândia         &     604,013 &  0.789 &     0.109 \\
Volta Redonda      &     579,427 &  0.747 &     0.086 \\
Feira de Santana   &     556,642 &  0.712 &     0.042 \\
Maringá            &     546,408 &  0.776 &     0.109 \\
Juiz de Fora       &     532,474 &  0.776 &     0.127 \\
Pelotas            &     525,503 &  0.741 &     0.091 \\
SJRP               &     502,494 &  0.789 &     0.128 \\
Macapá             &     499,466 &  0.725 &     0.068 \\
Caxias do Sul      &     499,199 &  0.781 &     0.096 \\
Petrolina-Juazeiro &     491,927 &  0.689 &     0.049 \\
Ipatinga           &     468,378 &  0.758 &     0.067 \\
Campos             &     463,731 &  0.716 &     0.046 \\
Campina Grande     &     452,162 &  0.705 &     0.071 \\
Crato              &     426,690 &  0.698 &     0.048 \\
Ilhéus-Itabuna     &     388,903 &  0.702 &     0.054 \\ \hline
\multicolumn{4}{l}{NH-SL: Novo Hamburgo/Sao Leopoldo} \\
\multicolumn{4}{l}{SJRP: Sao Jose do Rio Preto} \\
\multicolumn{4}{l}{SJC: Sao Jose dos Campos}
\end{tabular}
\end{table}
\end{document}